\documentclass{aa}
\usepackage{graphicx}
\usepackage{epsfig}
\usepackage{natbib}
\usepackage{graphics}
\newcommand{\be}{\begin{equation}}
\newcommand{\ee}{\end{equation}}
\newcommand{\source}{2FGL J2030.7+4417}
\newcommand{\hd}{HD 195592}

\begin{document}
\title{Is the bowshock of the runaway massive star HD 195592 a  {\it Fermi} source?}

   \author{M.~V. del Valle\inst{1,2,}\thanks{Fellow of CONICET, Argentina}
           G.~E. Romero,\inst{1,2,}\thanks{Member of CONICET, Argentina}
           \and M. De Becker\inst{3}}

   \offprints{Mar\'{\i}a V. del Valle : \\ {\em maria@iar-conicet.gov.ar}}
   \titlerunning{Gamma rays from the bowshock of a runaway star}

\authorrunning{del Valle et al. }  

\institute{Instituto Argentino de Radioastronom\'{\i}a, C.C.5, (1894) Villa Elisa, Buenos Aires,
Argentina. \and Facultad de Ciencias Astron\'omicas y Geof\'{\i}sicas,
Universidad Nacional de La Plata, Paseo del Bosque, 1900 La Plata, Argentina. \and Institut d'Astrophysique et de G\'eophysique, Universit\'e de Li\`{e}ge, 17, All\'ee du 6 Ao\^{u}t, B5c, B-4000 Sart Tilman, Belgium.}

\date{Received / Accepted}

% \abstract{}{}{}{}{}
% 5 {} token are mandatory

\abstract 
% context heatding (optional) % {} leave it empty if necessary 
{{\hd} is an O-type super-giant star, known as a well-established runaway. Recently, a {{\it Fermi}} $\gamma$-ray source ({2FGL J2030.7+4417}) with a position compatible with that of {\hd} has been reported. } 
{Our goal is to explore the scenario where {\hd} is the counterpart of the {{\it Fermi}} $\gamma$-ray source modeling the non-thermal emission produced in the bowshock of the runaway star.} 
{We calculate the spectral energy distribution of the radiation produced in the bowshock of {\hd} and we compare it with {{\it Fermi}} observations of the {\source}.}
{We present relativistic particle losses and the resulting radiation of the bowshock of {\hd} and show that the latter is compatible with the detected $\gamma$-ray emission.}
{We conclude that the {{\it Fermi}} source {\source} might be produced, under some energetic assumptions, by inverse Compton up-scattering of photons from the heated dust in the bowshock of the runaway star. {\hd} might therefore be the very first object detected belonging to the category of $\gamma$-ray emitting runaway massive stars, whose existence has been recently predicted.}


\maketitle

\section{Introduction}
The star {\hd} (DB+43\,3630) is a massive runaway visible from the northern hemisphere, located at a distance $\sim$ 1.1 kpc (see \citet{Schilbach2008}). There is strong evidence supporting the hypothesis that {\hd} is a binary system, with a period  of about 5 days \citep{DeBecker2010}. {\hd} produces a clearly detected bowshock, as it moves supersonically through the interstellar medium (ISM) as reported by \citet{Noriega1997}. 

\begin{figure}
\begin{centering}
\resizebox{.85\hsize}{!}{\includegraphics[trim=0cm 0cm 0cm 0cm, clip=true,angle=0]{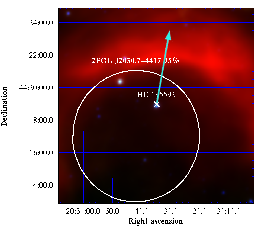}}
\caption{{\it WISE} (Wide-field Infrared Survey Explorer) RGB image at bands  W2 (4.6 $\mu$m) , W3 (12 $\mu$m)   and  W4 (22 $\mu$m) of the shocked gas around the runaway star {\hd}; the radiative heating of the swept-up dust produces the IR emission, and traces the bowshock. The 95 \% location error circle of the gamma-ray source {\source} is shown (the 99\% countour is outside the figure). The probability countours from gamma-ray sources are model$-$dependent and they must be taken as indicative only.}
\end{centering}
\label{wise}
\end{figure}

Relativistic particles can be accelerated at strong shocks produced by the stellar wind of a massive runaway interacting with the ISM (i.e. bowshocks). These particles can yield non-thermal radiation \citep{Benaglia2010, delValle2012}. Recently, a {\it Fermi} source, {\source}, that might be associated with {\hd}, has been reported in the Fermi Large Area Telescope Second Source Catalog \citep{Nolan2012}, see Fig. \ref{wise}. In this Letter, we apply the model developed by \citet{delValle2012} for the non-thermal emission that takes place in the bowshocks of runaway stars to {\hd}.  We confront then our theoretical spectral energy distribution with the measured flux of the {\it Fermi} source in order to explore the possibility of a physical association of the bowshock associated with {\hd} \citep[see][]{Peri2012} with {\source}. 

In Section\,\ref{star}, we make a census of the relevant information related to {\hd} with emphasis on previous observational results in different wavebands. Next, in Sect.\,\ref{rad}, we discuss the computation of the non-thermal emission and present the best-fit to the {\source} source. A brief discussion and our conclusions are given in Sect.\,\ref{end}.

\section{The stellar system {\hd}}\label{star}

{\hd} is an O9.5Ia-type runaway star that is thought to be originated in the open cluster NGC6913 \citep{Schilbach2008}. A detailed spectroscopic study revealed that {\hd} is a binary system with a period of a few days, with a lower mass early B-type companion \citep{DeBecker2010}. Low amplitude radial velocity variations were detected in every strong absorption line in the blue spectrum of {\hd}. The variations exhibit two time-scales:  $\sim$ 5.063 and $\sim$ 20 days. The 5.063-day variation is thought to be the period of the binary system associated with {\hd}. For the second time scale, \citet{DeBecker2010} give two possible explanations. It may be the signature of an additional star or it may be the signature of intrinsic variability related to the stellar rotation.

Previous radio observations with the Very Large Array (VLA) at 4.85, 8.45 and 14.95\,GHz revealed thermal emission with flux densities of a fraction of mJy, and a spectral index equal to 0.98 \citep{Scuderi1998}. Adopting a distance of 1.1\,kpc (\citet{Schilbach2008}), the integrated radio luminosity in the centimeter-domain is estimated to be of the order of 4\,$\times$\,10$^{28}$\,erg\,s$^{-1}$. It should be pointed out that the radio source reported by \citet{Scuderi1998} has an angular size of the order of a few arcseconds, and must therefore be associated with the stellar system itself and not with the more extended bowshock. If the bowshock  produces  radio emission, its flux level should be lower than the thermal contribution from the stellar winds of the O9.5 and B components, since \citet{Scuderi1998} did not report any extended non-thermal emission. 

In the soft X-ray domain, {\hd} was never target of a dedicated observation. However, the survey for point sources in the Cygnus region (including the position of {\hd}) performed by \citet{DeBecker2007} with the ISGRI instrument on-board the {\it INTEGRAL} satellite in hard X-rays allowed to derive upper limits for undetected point sources. Even though the background level in hard X-rays is not uniform, depending notably on the vicinity of bright X-ray sources, one could consider upper limits derived in regions of similar background level to be fairly applicable to the position of {\hd}\footnote{The upper limits published by \citet{DeBecker2007} were determined in a region located closer to the position of the bright X-ray source Cyg\,X-3 where the background level should be slightly higher than at the position of {\hd}. These values should therefore be considered as conservative. This does not affect their relevance in the context of this discussion.}. According to the flux upper limit values determined by \citet{DeBecker2007} and assuming once again a distance of 1.1 \,kpc, we estimate that our target should not be more luminous than about 7\,$\times$\,10$^{32}$ erg\,s$^{-1}$, 5\,$\times$\,10$^{33}$ erg\,s$^{-1}$ and 7\,$\times$\,10$^{33}$ erg\,s$^{-1}$, respectively, in the 20--60\,keV, 60--100\,keV and 100-1000\,keV energy bands. 

Finally, it might be worth commenting briefly on the potential role of binarity in the production of non-thermal radiation. It is indeed well established that at least some colliding-wind binaries are able to accelerate particles up to relativistic energies and consequently to produce non-thermal radiation \citep{BenRom2003,DeBeckerRev2007,BenagliaRev2010}. However, the short orbital period in {\hd} suggests that the stellar separation in the system would not allow relativistic electrons to reach Lorentz factors high enough as to produce a significant $\gamma$-ray emission as detected by Fermi. The particle acceleration process would indeed most probably be strongly inhibited by the strong ultraviolet/visible radiation fields from both stars through inverse Compton (IC) scattering, preventing the colliding-winds to emit significantly in the Fermi bandpass. In this paper, we will therefore explore the scenario where the $\gamma$-rays come from the bowshock produced by the stellar wind interacting with the ISM.

\section{Non-thermal emission calculation}\label{rad}

For the calculation of the non-thermal radiation we follow the model recently developed by \citet{delValle2012}.  The values for the relevant parameters are given in Table\,\ref{table}.

The collision of the supersonic stellar wind with the interstellar medium produces two shocks \citep[e.g.][]{Wilkin2000}. Relativistic particles are accelerated at the reverse shock, that propagates in the opposite direction of the motion of the star, inside the stellar wind. This shock is adiabatic and strong. The particle acceleration mechanism is diffusive first order shock acceleration \citep[e.g.][]{Bell1978}. The interactions of the locally injected relativistic particles with matter, radiation, and magnetic fields in the shocked wind produce non-thermal radiation by a variety of processes \citep{delValle2012}.  

The acceleration region is assumed to be a small region near the bowshock apex, of scale length $\sim$ $\Delta$, where $\Delta$ $\sim$ $M^{-2}R_{0}$. Here,  $M$ is the Mach number of the shocked wind and  $R_{0}$ is the so-called standoff radius \citep[e.g.][]{Wilkin1996}. In the case of {\hd}, we adopt $R_{0}$ $\sim$ 1.73 pc \citep{Peri2012}.

In order to roughly estimate the magnetic field in the flow, we assume that the magnetic energy density is in sub-equipartition with respect to the kinetic energy $L_{\rm T}$ of the wind\footnote{Otherwise the gas would be mechanically incompressible.}.  Therefore, we adopt the constraint $\chi < 1$. This means:
\begin{equation}
\frac{B^{2}}{8\pi} = \frac{ \chi L_{\rm T}}{V_{\rm w} A},
\label{equipar}
\end{equation}
where $A$ is the area of a sphere of radius $R_{0}$, and $L_{\rm T}$ is the available power in the system (best fits are provided by $\chi\sim 5\times 10^{-2}$). 

The kinetic power of the stellar wind is: 
\begin{equation}
L_{\rm T} \sim  \frac{1}{2}{\dot M_{\rm w}}V_{\rm w}^2. 
\end{equation}

For {\hd}, according to the best available data (Table \ref{table}): $L_{\rm T}$ $\sim$ 10$^{36}$ erg s$ ^{-1}$. Our estimate relies on the primary star parameters, even though we are dealing with a binary system. However, the contribution to the total kinetic power coming from the B-type component is expected to amount up to only a small fraction of that from the O super-giant, and would therefore not affect seriously our order of magnitude estimate.

The power available to accelerate particles in the reverse shock is $L = f L_{\rm T} \sim 2\times 10^{34}$ erg s$^{-1}$, where $f$ is the ratio of the volume of a sphere of
radius $R_{0}$ and the volume of the acceleration region. Some fraction of this power goes into relativistic particles $L_{\rm rel} = q_{\rm rel} L$. The energetics of the gamma-ray source requires $L_{\rm rel}$ $\sim$ $4\times 10^{33}$ erg s$^{-1}$, so  $q_{\rm rel}$ $\sim$ 0.2, which seems to be a reasonable value if compared, for instance, with supernovae, which are expected to convert about 20\% of their kinetic power into relativistic particles \citep[e.g.][]{Ginzburg1964}.

In the calculations of the spectral energy distribution we take into account both hadronic and leptonic content in the relativistic power, $L_{\rm rel} = L_{\rm p} + L_{\rm e}$. We consider $L_{\rm p} = L_{\rm e}$, which means equal efficiency in the acceleration of both types of particles. 

A proton-dominated scenario seems to be unlikely in the case of \hd . Since the fraction of the relativistic proton energy that goes to neutral pions in each interaction is of $\sim$ 17\% \citep[e.g.][]{Aharonian00}, in order to obtain the observed $\gamma$-ray luminosity a very high effciency in convenrting kinetic energy into relativistic particles is necessary. This would required very extended, or perhaps even multiple, acceleration sites in the bowsock. For simplicity, we stick here to the simplest hypothesis: equipatition beteween electrons and protons. This assumption, from the energetic point of view, is also the most conservative.

\begin{table}
\begin{center}
\caption[]{Parameters for HD 195592}
\begin{tabular}{lll}
\hline\noalign{\smallskip}
\multicolumn{2}{l}{Parameter} & value\\
\hline\noalign{\smallskip}
$R_{\rm 0}$ & Standoff radius & 1.73 pc \\
$\dot{M_{\rm w}}^{\rm (a)}$ & Wind mass loss rate& 3.3 $\times 10^{-7}$ M$_{\odot}$ yr$^{-1}$\\
$\alpha$ &Particle injection index & 2\\
$V_{\rm w}^{\rm (a)}$ & Wind velocity & 2.9$\times10^{8}$ cm s$^{-1}$ \\
${\chi}$ & Subequipartition factor & 5$\times10^{-2}$\\
$B$ & Magnetic field & $\sim 2\times 10^{-6}$ G  \\
$T_{\star}^{\rm (b)}$ & Star temperature & 2.8$\times 10^{4}$ K \\
$L_{\star}^{\rm (b)}$ & Star luminosity & 3.1$\times 10^{5}$ $L_{\odot}$ \\
$T_{\rm IR}$ & Dust temperature & $\sim$ 40  K\\
\hline\\
\end{tabular}	

{$^{\rm (a)}$ Values from \citet{Muijres2012}. 
$^{\rm (b)}$ Values from \citet{Mart2005}.}
 \label{table}
\end{center}
\end{table}

\begin{figure}
\begin{center}
\resizebox{.8\hsize}{!}{\includegraphics[trim=0cm 0cm 0cm 0cm, clip=true,angle=270]{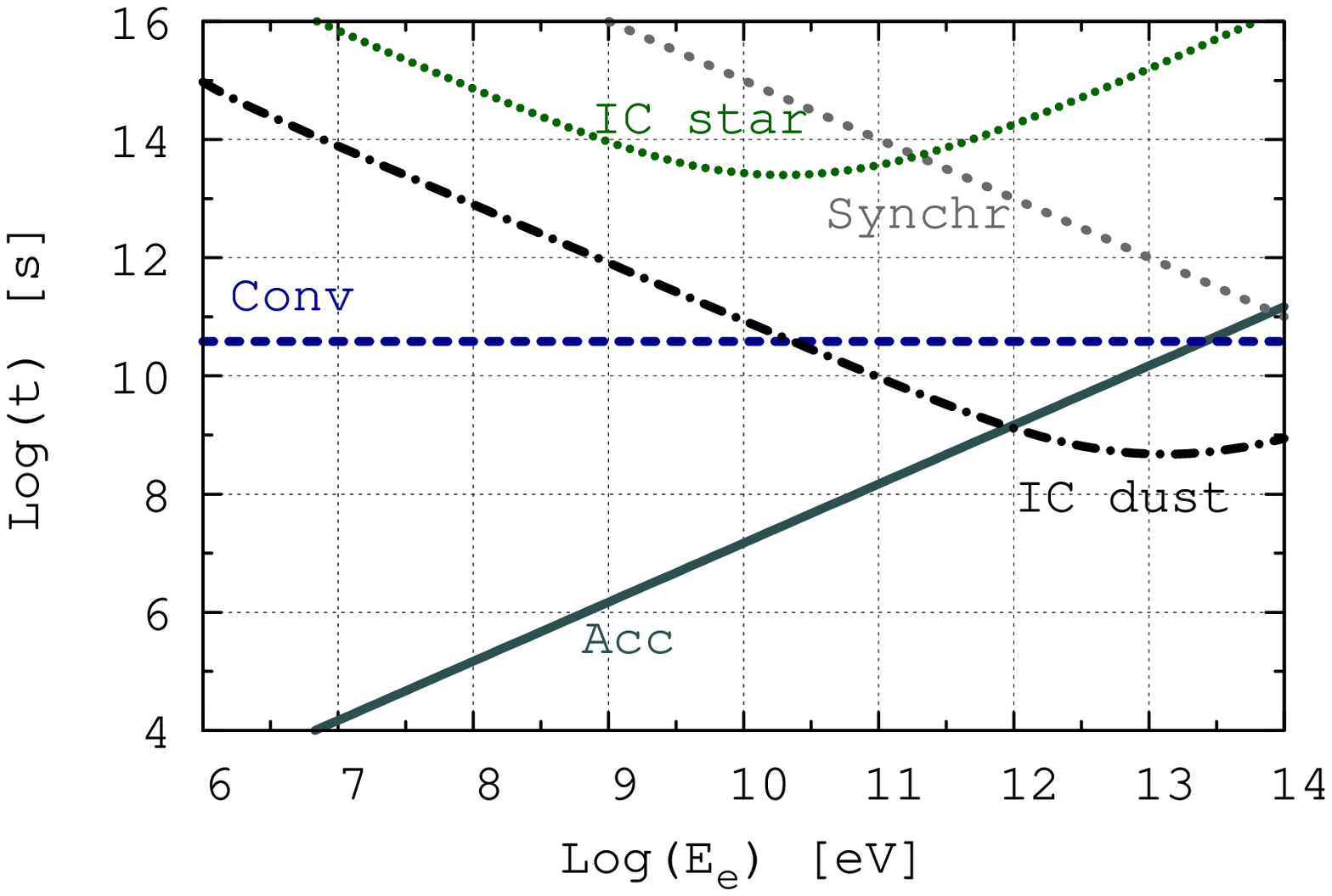}}
\resizebox{.8\hsize}{!}{\includegraphics[trim=0cm 0cm 0cm 0cm, clip=true,angle=270]{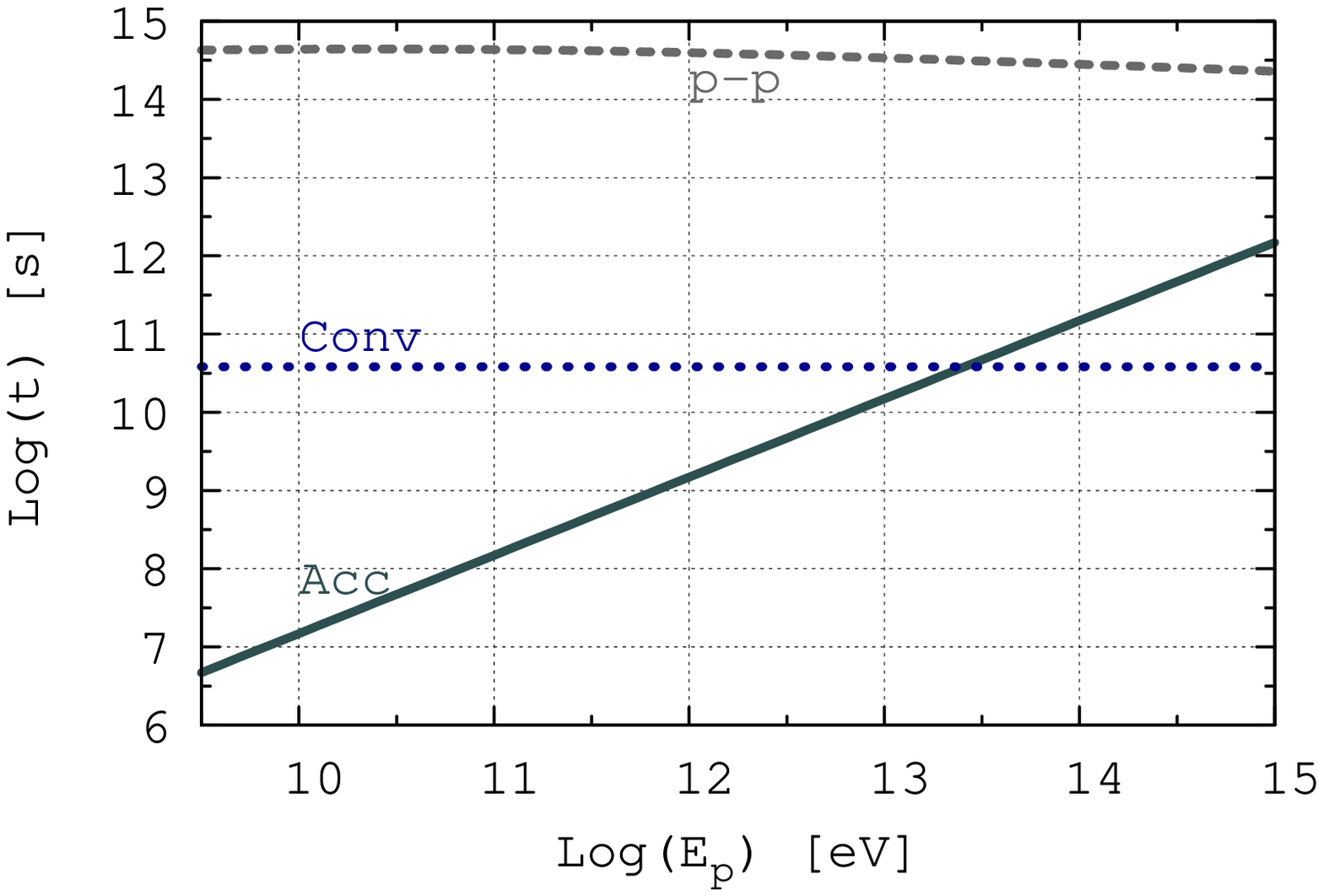}}
\caption{Acceleration and cooling time scales for electrons and protons for {\hd}. Up panel is for electrons and down panel is for protons.}
\label{fig:cool}
\end{center}
\end{figure}

The electrons lose energy mainly by IC scattering, synchrotron radiation, and 
relativistic Bremsstrahlung. Protons lose energy through proton-proton inelastic
collisions with the shocked wind material. The relativistic particles can escape from the
acceleration region convected away by the stellar wind. These non-radiative losses impose
the upper limit to the energy of protons. In Fig.\,\ref{fig:cool} we show the cooling
rates for both electrons and protons in the acceleration region. The IC scattering of IR
photons completely dominates the radiative losses.  Little power is radiated as
synchrotron radiation, rendering the non-thermal radio and X-ray counterparts quite weak 
in comparison to the IC $\gamma$-ray source.

As can be seen from Fig.\,\ref{fig:cool} the electrons reach energies $\sim 0.6$  TeV while 
the protons can get $\sim$ $15$ TeV. In both cases the Hillas criterion is satisfied, i.e. 
$E_{\rm max} < 300 ({\Delta}R_{0}/{\rm cm}) (B/{\rm G})$ eV.

To determine the steady state particle distributions for electrons and protons, we solved the transport equation in steady state \citep{Ginzburg1964}. We refer to \citet{delValle2012} for details of calculation. Although most protons are convected away before cooling -- see Fig.\ref{fig:cool} -- downstream instabilities can produce some mixing with external gas. We consider that a fraction $f_{p}$ ($\sim$ 10\%) of the convected protons interact downstream with the gas, and hence enhance proton-proton contribution. Even if all protons were to cool, the hadronic contribution would remain minor, except at energies above 1 TeV. 
At the energies of interest in this paper both the emission from the shocked ISM and the absorption are negligible.

\begin{figure}
\begin{center}

\includegraphics[width=0.7\linewidth,angle=270]{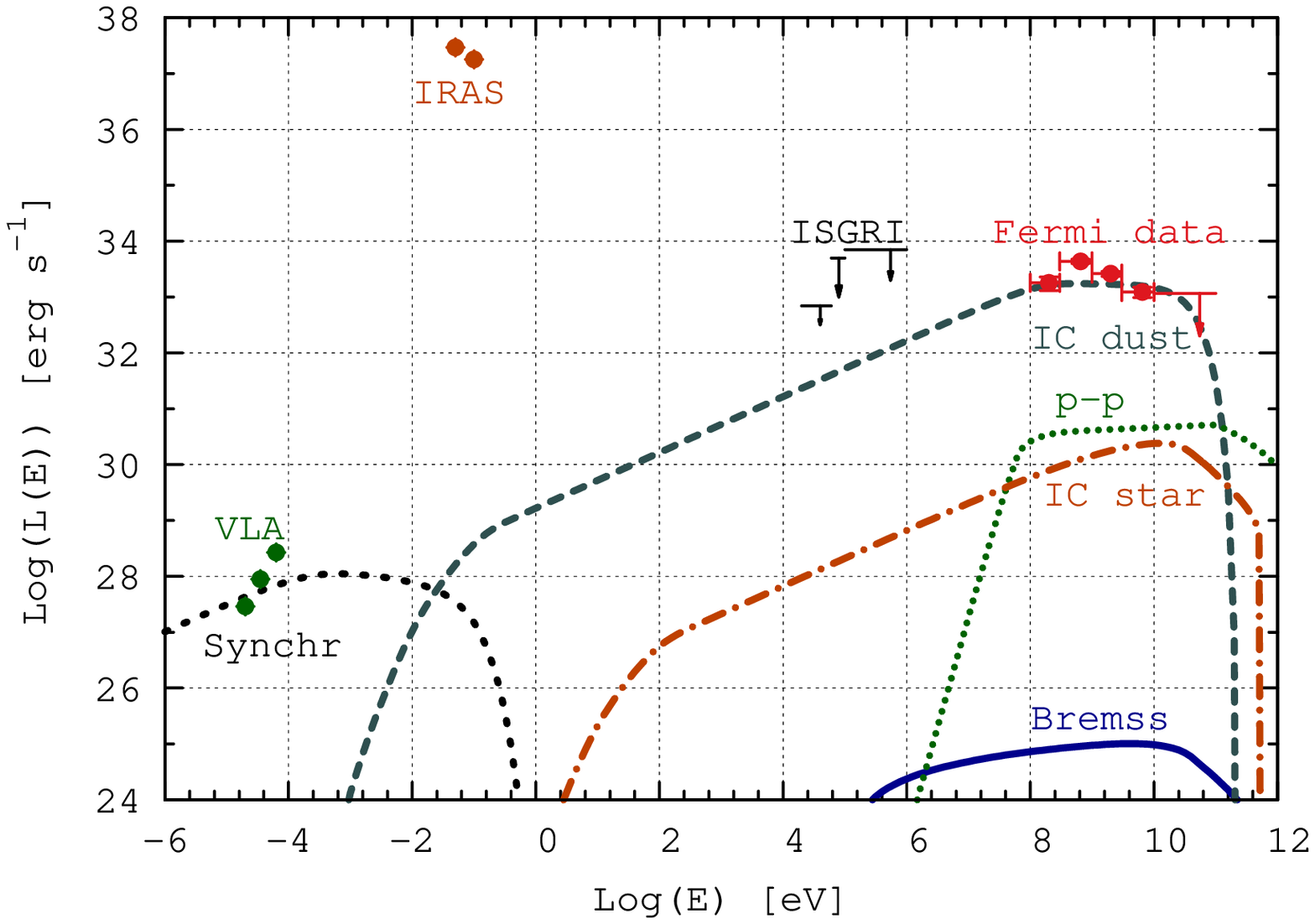}
\caption{Computed SED for {\hd} bowshock,  at d $\sim$ 1.1 kpc and  {\it Fermi} data of {\source}. The hard X-rays upper limits, thermal radio data (VLA) (\citealt{Scuderi1998}) and the IR --IRAS-- emission (\citealt{VanBuren1995}) are also shown.}
\label{SEDz}
\end{center}
\end{figure}

Figure\,\ref{SEDz} shows the computed spectral energy distribution (SED) for the emission produced at the bowshock of {\hd}. {\it Fermi} data are also shown along with data at other wavelengths.  The IR emission from {\it IRAS} is produced by the heated dust  \citep{VanBuren1995}, as the {\it WISE} emission. Only the $\gamma$-ray flux is non-thermal, while at other wavelength the emission is mainly thermal. In the calculation we have taken into account only the wind of the primary star of the potential binary system, so our estimates can be considered as energetically conservative. 

The SED derived by our calculation lends support to the scenario where the $\gamma$-ray emission from {\source} could come from the bowshock produced by {\hd}. Our SED is also compatible with the lack of detection of any significant hard X-rays with INTEGRAL at the same position. In addition, the predicted synchrotron radio flux is  too low to have been detected by previous radio investigations in the vicinity of {\hd}, where the thermal radio emission dominates.  

\section{Discussion and conclusions}\label{end}

Massive stars with strong winds are suspected to be $\gamma$-ray sources since the early 1980s \citep{Casse1980,Voelk1982,White1991,White1992}. Despite some statistical evidence \citep{Montmerle1979,Romero1999}, conclusive identifications remain elusive to this day. This is not surprising taking into account the strong non-radiative losses experienced by relativistic particles in the stellar winds \citep{Voelk1982} and the strong absorption expected close to the massive star \citep{Romero2010}. The best prospect, then, is the detection of high-energy photons wherever strong shocks can re-accelerate electrons and ions far from the star. This is the case of combined effects of massive stars in stellar associations \citep{Torres2004} and colliding wind binaries \citep{Eichler1993,BenRom2003,DeBeckerRev2007}. Recent detections of Westerlund 2 \citep{Aharonian2007} and Eta-Carina \citep{Tavani2009,Abdo2010}, at TeV and GeV $\gamma$ rays, respectively, seem to support this picture. 

Runaway massive stars offer a unique opportunity to detect GeV-TeV emission from single massive stars. The stagnation point of the wind of these stars is located at sufficient distance as to preclude, under the adequate viewing angles, significant $\gamma$-ray absorption. The recent detection of non-thermal radio emission from the bowshock of BD +43$^{\circ}$3654 by \citet{Benaglia2010} and  the non-thermal X-ray emission reported recently from AE Aurigae \citet{Javier2012} confirms the capability of some of these stars to accelerate at least electrons up to relativistic energies. The presence of rich infrared photon fields locally generated by the heated dust swept by the shocks guarantees suitable targets for IC interactions, that might yield, in some cases, detectable $\gamma$-ray fluxes.     

The star {\hd} presents some characteristics (e.g. strong IR field, distant stagnation point due to the relatively small stellar velocity in a dense medium, absence of any other source in the {\it Fermi} location error box) that makes it a good candidate to be the very first $\gamma$-ray emitting bowshock runaway identified so far. A confirmation of the nature of this source would require deep X-ray observations to check whether there is a power-law spectrum as expected from our modeling. {\hd} is therefore a good candidate for additional observations for instance, with ASTRO-H (JAXA mission to be launched in 2014, \citealt{Takahashi2010}) to investigate non-thermal hard X-rays, and with ACIS on the {\it Chandra} X-ray Observatory because of its low background and high spatial resolution necessary to spatially disentangle the soft thermal emission from the binary and the expected soft non-thermal X-rays from the bowshock.

\begin{acknowledgements}
We thank an anonymous referee for constructive comments. This work is supported by PIP 0078 (CONICET) and PICT 2007-00848, Pr\'estamo BID (ANPCyT). GER received additional support from the Spanish Ministerio de Inovaci\'on y Tecnolog\'{\i}a  under grant AYA 2010-21782-c03-01. We thank P. Benaglia for help with the IR data. 
%The SIMBAD database was used for the bibliography. This publication makes use of data products from the Wide-field Infrared Survey Explorer, which is a joint project of the University of California, Los Angeles, and the Jet Propulsion Laboratory/California Institute of Technology, funded by the National Aeronautics and Space Administration.
\end{acknowledgements}

\end{document}